\documentstyle[emulateapj]{article}
 
\font\smcap=cmcsc10

\newcommand\hi{{\smcap H$\,$i}}
\newcommand\Ha{H$\alpha$}
\newcommand\HaN{H$\alpha$ + [{\smcap N$\,$ii}]}
\newcommand\NHI{$N_{HI}$}
\newcommand\Myr{$M_\odot \, {\rm yr}^{-1}$}

\newcommand\mR{$\mu_R$}
\newcommand\msqas{mag arcsec$^{-2}$}
\newcommand\B{$B$}

\newcommand\R{$R$}
\newcommand\about{$\sim$}
\newcommand\as{\arcsec}
\newcommand\kps{km s$^{-1}$}
\newcommand\ie{{\it i.e.}}
\newcommand\et{{\it et~al.}}
\def\col#1{$\times 10^{#1} {\rm cm}^{-2}$}

\lefthead{Hibbard, Vacca, \& Yun}
\righthead{HI and Stellar Tidal Displacements}
\submitted{To appear in the March 2000 issue of The Astronomical Journal}

\begin{document}

\title{The Neutral Hydrogen Distribution in Merging Galaxies:\\
Differences between Stellar and Gaseous Tidal Morphologies} 

\author{J. E.~Hibbard\altaffilmark{1}}
\affil{National Radio Astronomy Observatory\altaffilmark{2}, 520 Edgemont 
Road, Charlottesville, VA, 22903; jhibbard@nrao.edu}

\author{W. D.~Vacca} 
\affil{IRTF, Institute for Astronomy, 2680 Woodlawn Drive, 
Honolulu, HI 96822; vacca@ifa.hawaii.edu}

\author{M. S.~Yun}
\affil{National Radio Astronomy Observatory\altaffilmark{2}, P.O.\   Box 0, 
Socorro, New Mexico, 87801; myun@nrao.edu}

\bigskip

\altaffiltext{1}{Visiting Astronomer, Kitt Peak National
Observatory, National Optical Astronomy Observatories, which is
operated by the Association of Universities for Research in Astronomy,
Inc. (AURA) under cooperative agreement with the National Science
Foundation.}
\altaffiltext{2}{The National Radio Astronomy Observatory is 
operated by Associated Universities, Inc., under cooperative agreement 
with the National Science Foundation.}

\begin{abstract}

As part of several \hi\ synthesis mapping studies of merging galaxies,
we have mapped the tidal gas in the three disk-disk merger systems Arp
157 (NGC 520), Arp 220, and Arp 299 (NGC 3690).  These systems differ
from the majority of the mergers mapped in \hi, in that their stellar
and gaseous tidal features do not coincide. In particular, they
exhibit large stellar tidal features with little if any accompanying
neutral gas and large gas-rich tidal features with little if any
accompanying starlight. On smaller scales, there are striking
anti-correlations where the gaseous and stellar tidal features appear
to cross.  We explore several possible causes for these
differences, including dust obscuration, ram pressure stripping, and
ionization effects.  No single explanation can account for all of the
observed differences.  The fact that each of these systems shows
evidence for a starburst driven superwind expanding in the direction
of the most striking anti-correlations leads us to suggest that 
the superwind is primarily responsible for the observed differences, 
either by sweeping the features clear of gas via ram pressure, or by
excavating a clear sightline towards the starburst and allowing UV
photons to ionize regions of the tails. If this suggestion is correct,
only systems hosting a galactic superwind and experiencing a
high-inclination encounter geometry (such that tidal gas is lifted
high above the starburst regions) should exhibit such extreme
differences between their \hi\ and optical tidal morphologies.

\end{abstract}


\section{Introduction}
\label{sec:intro}

Nearly 30 years ago, Toomre \& Toomre (1972) elegantly demonstrated that
the tails and bridges emanating from many peculiar galaxies may arise
kinematically from dynamically cold disk material torn off of the outer
regions of galaxies experiencing strong gravitational interactions. 
Early spectroscopic studies of gas within the tidal tails of merging
galaxies provided observational support for this hypothesis by showing
the tails to have the kinematics expected for a gravitational origin
(e.g.~Stockton 1974a,b).  \hi\   mapping studies are particularly well
suited to such studies, as the tidally ejected disk material is usually
rich in neutral hydrogen and can be traced to very large distances from
the merging systems (e.g.~van der Hulst 1979; Simkin \et\   1986; Appleton
\et\   1981, 1987; Yun \et\   1994).  Once mapped, the tidal kinematics can
be used either alone, to disentangle the approximate spin geometry of
the encounter (Stockton 1974a,b; Mihos \et\   1993; Hibbard \& van Gorkom
1996, hereafter HvG96; Mihos \& Bothun 1998), or in concert with
detailed numerical models, to constrain the full encounter geometry
(e.g.~Combes 1978; Combes \et\   1988; Yun 1992, 1997; Hibbard \& Mihos
1995; Gardiner \& Noguchi 1996). 

However, not all systems can be easily explained by purely
gravitational models such as those used by Toomre \& Toomre. For
example, gravitational forces by themselves should not lead to
differences between stellar and gaseous tidal components.  Numerical
models which include hydrodynamical effects do predict a
decoupling of the dissipative gaseous and non-dissipative stellar
components (e.g.~Noguchi 1988; Barnes \& Hernquist 1991, 1996; Weil \&
Hernquist 1993; Mihos \& Hernquist 1996; Appleton, Charmandaris \&
Struck 1996; Struck 1997), but only in the inner regions or along
bridges where gas orbits may physically intersect (see e.g.~Fig.~4 of
Mihos \& Hernquist 1996).  Decoupling of the gaseous and stellar
components within the tidal tails is not expected.

Nonetheless, differences between the optical and gaseous tidal
morphologies have been observed.  These differences can be subtle,
with the peak optical and \hi\   surface brightnesses simply displaced
by a few kpc within the tails (e.g.  NGC 4747, Wevers \et\   1984; NGC
2782 Smith 1994; NGC 7714/4 Smith \et\   1997; Arp 295A, NGC 4676B, and
NGC 520 Southern tail, Hibbard 1995, HvG96), or they can be extreme,
with extensive \hi\   tidal features apparently decoupled
from, or even anti-correlated with, the optical tidal features.  It is
this latter category of objects that we wish to address in this
paper. In particular, we address the morphology of the tidal gas
and starlight in the merging systems NGC 520 (Arp 157), Arp 220, and
Arp 299 (NGC 3690). 

The three systems were observed as part of our on-going studies on the
tidal morphologies of optically and IR selected mergers (Hibbard 1995,
HvG96, Hibbard \& Yun 1996 and in prep.).  These studies involve
moderate resolution ($\theta_{FWHM}\sim 15^{\prime\prime}$) VLA~\hi\
spectral-line mapping observations and deep optical \B\   and \R\
broad-band imaging with large format CCDs using the KPNO 0.9m (NGC 520)
and the University of Hawaii 88\as\   telescopes.  The \hi\   and optical
observations, reduction, and data products have been presented in
Hibbard (1995) and HvG96 for NGC 520, in Hibbard \& Yun (1999, hereafter
HY99) for Arp 299, and in Yun \& Hibbard (2000; see also Hibbard \& Yun
1996) for Arp 220.  We refer the reader to these papers for details of
the observations and data reduction.  These systems are extremely
disturbed, and we cannot hope to offer a full description of their
peculiarities here.  For more information we refer the reader to the
above references. 

\section{Observed Stellar and Gaseous Tidal Morphologies}

Figures~\ref{fig:n520mos}--\ref{fig:a220mos} show the optical and
atomic gas morphologies of each of the three systems discussed here.
For NGC 520 and Arp 220 only the inner regions are shown in order to
highlight the differences we wish to address.  Panel {\bf (a)} presents 
a greyscale representation of the optical morphology of each system 
with features of interest labeled.  Panel {\bf (b)} shows
the \hi\  distribution.  Contours indicate the distribution of \hi\
mapped at low-resolution ($\theta_{FWHM}\sim 30^{\prime\prime}$),
whereas the greyscales show the \hi\  mapped at higher resolution
($\theta_{FWHM}\sim 15^{\prime\prime}$).  The former is sensitive to
diffuse low column density (\NHI) neutral hydrogen, while the latter
delineates the distribution of the higher column density \hi.  The
central region of each \hi\  map appears to have a hole (indicated by
the dotted contours), which is due to \hi\  absorption against the
radio continuum associated with the disk-wide starbursts taking place
in each galaxy (see Condon \et\  1990).  In panel {\bf (c)}, we again
present the optical morphology in greyscales, and the higher
resolution \hi\  distribution as contours.  Finally, panel {\bf (d)}
presents a smoothed, star-subtracted \R-band image contoured upon a
greyscale representation of the high-resolution \hi\  map.

In the final panels of Figs.~\ref{fig:n520mos}--\ref{fig:a220mos}
dashed lines labeled ``Slice'' indicate the locations from which \hi\
and optical intensity profiles have been extracted; these profiles are
plotted in Figure~\ref{fig:slices}.  Arrows labeled ``Superwind"
indicate the position angle (P.A.) of \Ha\ or soft x-ray plumes,
believed to arise from a starburst-driven outflow or galactic
superwind in each system.  Such outflows are common in other IR bright
starbursts (e.g.~Heckman, Armus \& Miley 1987, 1990 hereafter HAM90;
Armus, Heckman, \& Miley 1990; Lehnert \& Heckman 1996), and are
thought to arise when the mechanical energy from massive stars and
supernovae in the central starburst is sufficient to drive the dense
interstellar medium outward along the minor axis (e.g.~Chevalier \&
Clegg 1985; Joseph \& Wright 1985; Suchkov \et\ 1994).  Often, such
starbursts are powerful enough to drive a freely expanding wind of hot
plasma completely out of the galaxy (``blowout"; HAM90).

In the following subsections we briefly discuss what is known about
the dynamical state of each system, and describe the differences
between the stellar and gaseous tidal morphologies.  Throughout this
paper distances and other physical properties are calculated assuming
$H_0 = 75\,~\rm km\,s^{-1}\,Mpc^{-1}$.

\subsection{NGC 520}
\label{sec:n520}

\begin{figure*}[t]
\plotfiddle{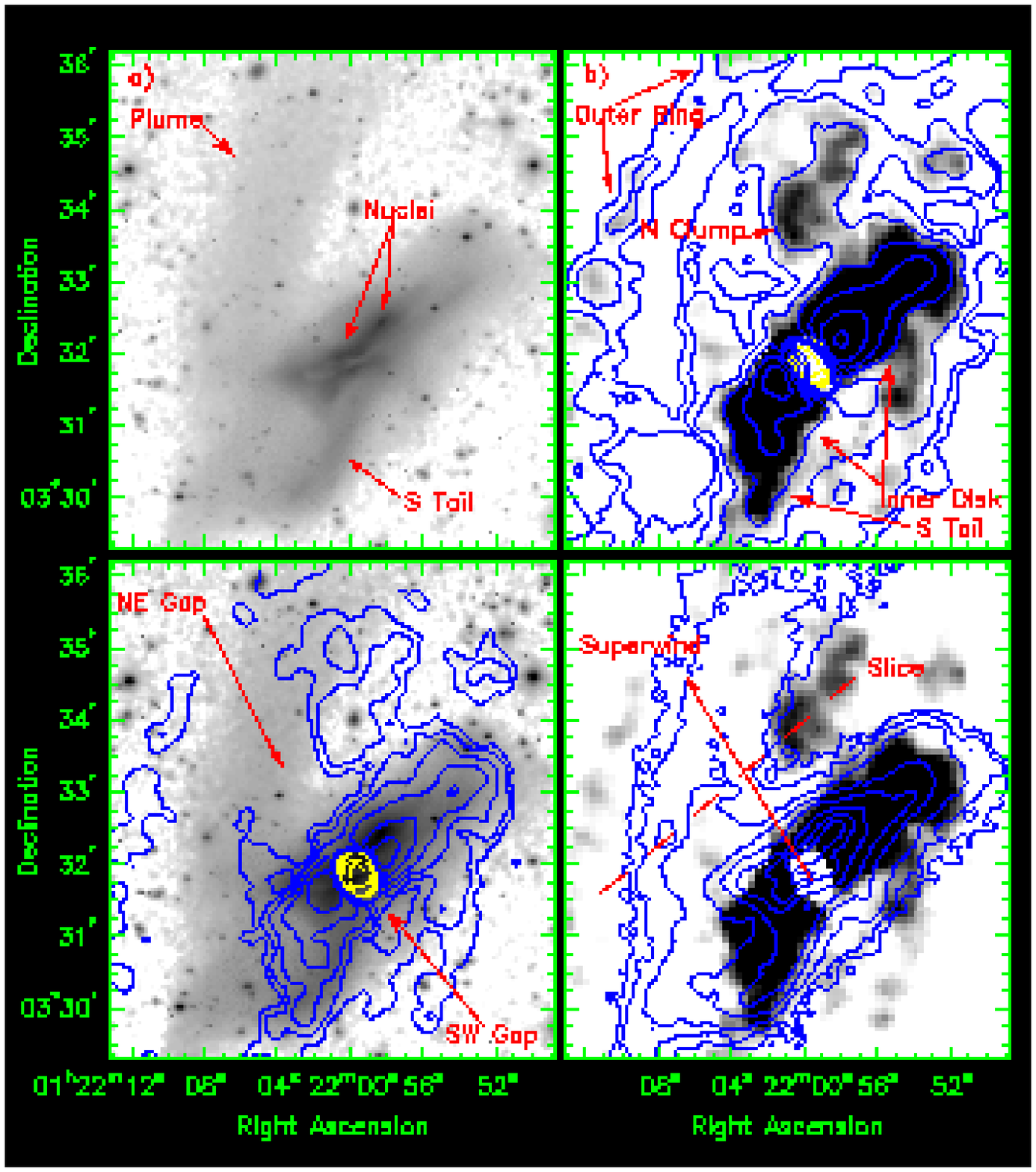}{17cm}{0}{80}{80}{-250}{-90}
\caption{{\sc HI and Optical Morphology of NGC 520.} \small
Panel (a) (top left)
displays an \R-band image with features of interest labeled. Panel (b)
(upper right) presents a greyscale and contour representation of the 
\hi\  data. The contours indicate the distribution of \hi\  mapped at 
low-resolution ($\theta_{FWHM}\sim 25^{\prime\prime}$), whereas the
greyscales show the \hi\  mapped at higher resolution ($\theta_{FWHM}
\sim 17^{\prime\prime}$). The lowest contour is drawn at a column 
density of 5\col{19}, with successive contours a factor of 2 higher.
The greyscales range from 1\col{20} (white) to 4\col{20} (black). 
Panel (c) (lower left) presents the optical image in 
greyscales with contours from the 
high-resolution \hi\  data superimposed. The lowest contour is drawn 
at a column density of 1\col{20}, with successive contours a factor 
of 2 higher.  Panel (d) (lower right) shows contours from a
star-subtracted \R-band image (lowest contour of 27 \msqas\  and contour
interval of 1 \msqas) upon greyscales of the high-resolution \hi\
data (from 1\col{20} to 4\col{20}). The dashed line labeled ``Slice"
indicates the locations of the intensity profiles plotted in
Fig.~\ref{fig:slices}a. The arrow labeled ``Superwind" indicates the 
direction (but not the extent) of the putative expanding superwind.}
\label{fig:n520mos}
\end{figure*}

NGC 520 (Arp 157, UGC 966) is an intermediate-stage merger, with the
two progenitor nuclei separated by 40\as\   (5.8 kpc for $D$= 30 Mpc,
1\as\   = 145 pc) and embedded within a common luminous envelope (HvG96
and references therein; see Fig.~\ref{fig:n520mos}a).  There is a
bright optical tidal tail stretching 24 kpc to the southeast
(henceforth referred to as the S Tail) which bends sharply eastward
and connects onto a broad optical plume\footnote{We follow the naming
convention of Schombert \et\   1990 and refer to tidal features with 
flat intensity profiles as plumes, and ones with Gaussian
profiles as tails}.  This plume continues to the north and west for 60
kpc before it appears to connect onto extended light surrounding
the dwarf galaxy UGC 957 (outside the region plotted in
Fig.~\ref{fig:n520mos}a; see Stockton \& Bertola 1980).

The primary nucleus (the easternmost nucleus in Fig.~\ref{fig:n520mos}a)
possesses a massive ($\sim 5\times 10^{9} M_\odot$) 1 kpc-scale rotating
molecular gas disk (Sanders \et\   1988; Yun \& Hibbard 1999).  An \hi\
disk is kinematically centered on this molecular disk, and extends to a
radius of $\sim$ 20 kpc (labeled ``Inner Disk'' in
Fig.~\ref{fig:n520mos}b).  Beyond this there is an intermediate ring of
\hi\   with a mean radius of $\sim$ 30 kpc (i.e., the material which
contains the feature labeled ``N Clump'' in Fig.~\ref{fig:n520mos}b),
and a nearly complete outer ring of \hi\  with a mean radius of 60 kpc,
which extends smoothly through the dwarf galaxy\footnote{The HI
kinematics show that UGC~957 is kinematically associated with this
outer gas ring, although it may lie slightly above or below it.
Rudimentary numerical modeling has shown that it is unlikely that the
UCG galaxy is responsible for the main optical features of the NGC~520
system (Stanford \& Balcells 1991).  It is unclear whether this system
is an interloper or was recently assembled from the surrounding ring
of gas.} UCG 957 (only partially seen in Fig.~\ref{fig:n520mos}b). 
There is a kinematic and morphological continuity between the molecular 
gas disk, the inner \hi\  disk, and outer \hi\  ring (Yun \& Hibbard 
1999), which suggests that all of this material is associated with the 
primary nucleus.  

The observations suggests that the NGC 520 interaction involved a
prograde-retrograde or prograde-polar spin geometry: The linear
morphology of the optical tail-to-plume system is typical of features
produced by a disk experiencing a prograde encounter (i.e.,
the disk rotates in the same direction as the merging systems orbit
each other).  The disk-like morphology and rotational kinematics of
the large-scale HI and the lack of any aligned linear tidal features,
on the other hand, are more typical of polar or retrograde encounter
geometries (i.e., disk rotation either perpendicular to or opposite
the direction of orbital motion). Such encounters fail to raise
significant tails (Toomre \& Toomre 1972; Barnes 1988), and much of
the disk material remains close to its original rotational plane. 

Neither the intermediate nor the outer \hi\   ring has an optical
counterpart (\mR $>$ 27 \msqas).  Despite smooth rotational kinematics,
this outer \hi\   has a very clumpy and irregular morphology, with notable
gaps near the optical minor axis (labeled ``NE gap'' and ``SW gap'' in
Fig.~\ref{fig:n520mos}c).  This figure shows that the outer \hi\   and
optical structures are anti-correlated, with the peak \hi\   column
densities (associated with the N clump) located to one side of the
optical plume.  In Fig.~\ref{fig:n520mos}d the \hi\   clump appears to be
bounded on three sides by the optical contours.  In Fig.~\ref{fig:slices}a 
we present an intensity profile at the location indicated by the dotted
line in Fig.~\ref{fig:n520mos}d, showing that the gas column density 
increases precisely where the optical light decreases. 

     While the \hi\ features exhibit a clear rotational kinematic
     signature, the well-defined edges and nearly-linear structure of the
     optical plume suggests that its constituent stars are moving
     predominantly along the plume, rather than in the plane of the sky:
     any substantial differential rotation would increase the width of the
     plume and result in a more disk-like morphology.  We therefore
     conclude that the gas rings and optical plume are both morphologically
     and kinematically distinct entities.  
This suggests that the observed 
gas/star anti-correlation is either transient (and fortuitous) or 
actively maintained by some process.

A deep \Ha\ image of NGC 520 shows plumes of ionized gas emerging both
north and south along the minor axis and reaching a projected height
of 3 kpc from the nucleus (HvG96).  It has been suggested that this
plume represents a starburst-driven outflow of ionized gas (HvG96,
Norman \et\ 1996).  The position angle of this plume is indicated by
an arrow in Fig.~\ref{fig:slices}d (P.A.  = 25\fdg).  This direction
corresponds to the most dramatic \hi/optical anti-correlations
mentioned above, and in the following we suggest that this region of
the optical tail actually lies directly in the path of the out-flowing
wind.

\subsection{Arp 299}
\label{sec:a299}

\begin{figure*}[t]
\plotfiddle{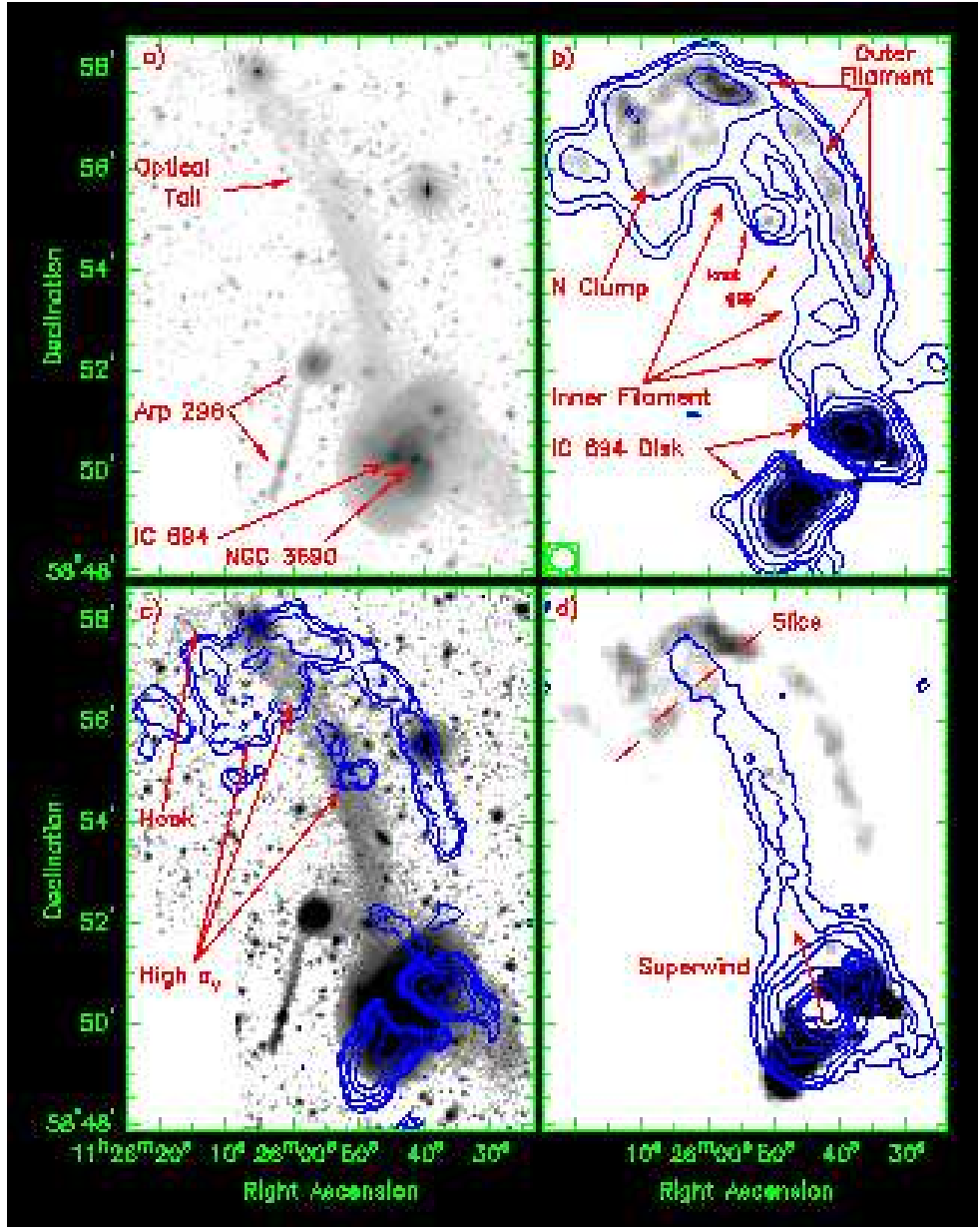}{18cm}{0}{70}{70}{-220}{-30}
\caption{{\sc HI and Optical Morphology of Arp 299.} \small
Panel (a) (top left)
displays an \R-band image with features of interest labeled. Panel (b)
(upper right) presents a greyscale and contour representation of the 
\hi\  data. The contours indicate the distribution of \hi\  mapped at 
low-resolution ($\theta_{FWHM}\sim 35^{\prime\prime}$), whereas the
greyscales show the \hi\  mapped at higher resolution ($\theta_{FWHM}
\sim 20^{\prime\prime}$). The lowest contour is drawn at a column 
density of 2.5\col{19}, with successive contours a factor of 2 higher.
The greyscales range from 1\col{20} (white) to 4\col{20} (black). 
Panel (c) (lower left) presents the optical image in greyscales 
with contours from the 
high-resolution \hi\  data superimposed. The lowest contour is drawn 
at a column density of 5\col{19}, with successive contours a factor 
of 2 higher.  Panel (d) (lower right) shows contours from a
star-subtracted \R-band image (lowest contour of 27 \msqas\  and contour
interval of 1 \msqas) upon greyscales of the high-resolution \hi\
data (from 1\col{20} to 4\col{20}). The dashed line labeled ``Slice"
indicates the locations of the intensity profiles plotted in
Fig.~\ref{fig:slices}b. The arrow labeled ``Superwind" indicates the 
direction (but not the extent) of the putative expanding superwind.}
\label{fig:a299mos}
\end{figure*}

Arp 299 (NGC 3690/IC 694, UGC 6471/2, Mrk 171, VV 118) is also an
intermediate stage merger, with two disk systems (IC 694 to the east,
NGC 3690 to the west --- see Figure~\ref{fig:a299mos}a) in close
contact but with their respective nuclei separated by 20\as\ (4.7 kpc
for $D$= 48 Mpc, 1\as\ = 233 pc).  A long, narrow, faint (\mR$\gtrsim$
26\msqas) tidal tail stretches to the north to a radius of $\sim$ 125
kpc.  \hi\ imaging of this system by HY99 (see also Nordgren et
al.~1997) shows a rotating gas-rich disk within the inner regions, and
a pair of parallel \hi\ filaments extending to the north.  From the
\hi\   morphology and kinematics, HY99 deduce that Arp 299 is the result
of a prograde-retrograde or prograde-polar encounter between two
late-type spirals, with the inner \hi\   disk associated with the
retrograde disk of IC~694, and the northern optical tail and tidal \hi\
filaments ejected by the prograde disk of NGC~3690. 

The parallel-filament or bifurcated morphology of the tidal \hi\  is
quite unlike that of the optical tail.  The inner \hi\  filament (so
labeled in Fig.~\ref{fig:a299mos}b) is of lower characteristic column
density (\NHI$\lesssim$ 8\col{19}) and is associated with the low surface
brightness stellar tail (Fig.~\ref{fig:a299mos}c).  The gas in this
filament has a more irregular morphology than that in the outer
filament (e.g., the ``gap'' and ``knot'' in Fig.~\ref{fig:a299mos}b),
and much of this material is detectable only after a substantial
smoothing of the data (Fig.~\ref{fig:a299mos}b).  The outer filament
is characterized by a higher \hi\  column density (\NHI\about
1.5\col{20}) but has no optical counterpart (\mR $>$ 27.5 \msqas).
This filament is displaced by approximately 20 kpc (in projection) to
the west of the inner filament for much its length, after which the
filaments merge together in a feature labeled the ``N Clump'' in
Fig.~\ref{fig:a299mos}b.  The parallel filaments have nearly identical
kinematics along their entire lengths, and join smoothly at the N
Clump.  This implies that these features form a single physical
structure.  

Based on preliminary numerical simulations, HY99 suggest that a 
bifurcated morphology can arise quite naturally during tail formation.
This occurs when the optically faint, gas-rich outer regions of 
the progenitor disk are projected adjacent to optically brighter regions 
coming from smaller initial radii\footnote{To produce filaments as well
separated as found in Arp 299 HY99 suggest that the bifurcation is
exacerbated by a pre-exisiting gaseous warp in the progenitor disk.}
(see also Mihos, 2000). However, this scenario does not explain 
why the inner filament, presumably drawn from optically bright but
still gas-rich material within the optical disk of the progenitor, 
should lack accompanying \hi. 

As in NGC 520, there is an anti-correlation between the optical and
gaseous column densities across the N clump, with the highest gas column
densities (2--3\col{20}) located on either side of the optical tail. 
This is illustrated in Fig.~\ref{fig:slices}b, where we plot a profile
along the position indicated by the dotted line in
Fig.~\ref{fig:a299mos}d.  The optical tail emerges above the N clump,
and appears to curve exactly around the northern edge of the N clump
(labeled ``hook'' in Fig.~\ref{fig:a299mos}c).  Also labeled in
Fig.~\ref{fig:a299mos}c are the three regions with anomalously high \hi\
velocity dispersions noted by HY99 ($\sigma_{HI} \sim$ 13--20 \kps\,
compared with $\sigma_{HI}\sim$ 7--10 \kps\   for the remainder of the
tail; see Fig.~7d of HY99); we will will refer to these regions in the
discussion (\S~\ref{sec:RPSwind}). 

Within the main body of Arp 299, vigorous star formation is taking
place, with an inferred star formation rate (SFR) of 50 \Myr\   (HAM90).
Recent X-ray observations reported by Heckman \et\   (1999) show
evidence for hot gas emerging from the inner regions and reaching 25
kpc to the north, which the authors interpret as evidence for a hot,
expanding superwind.  The position angle of this feature
(P.A.=25\fdg) is indicated by the arrow in Fig.~\ref{fig:a299mos}d, and
points towards the inner tidal filament and N clump.

\subsection{Arp 220}
\label{sec:a220}

\begin{figure*}[t]
\plotfiddle{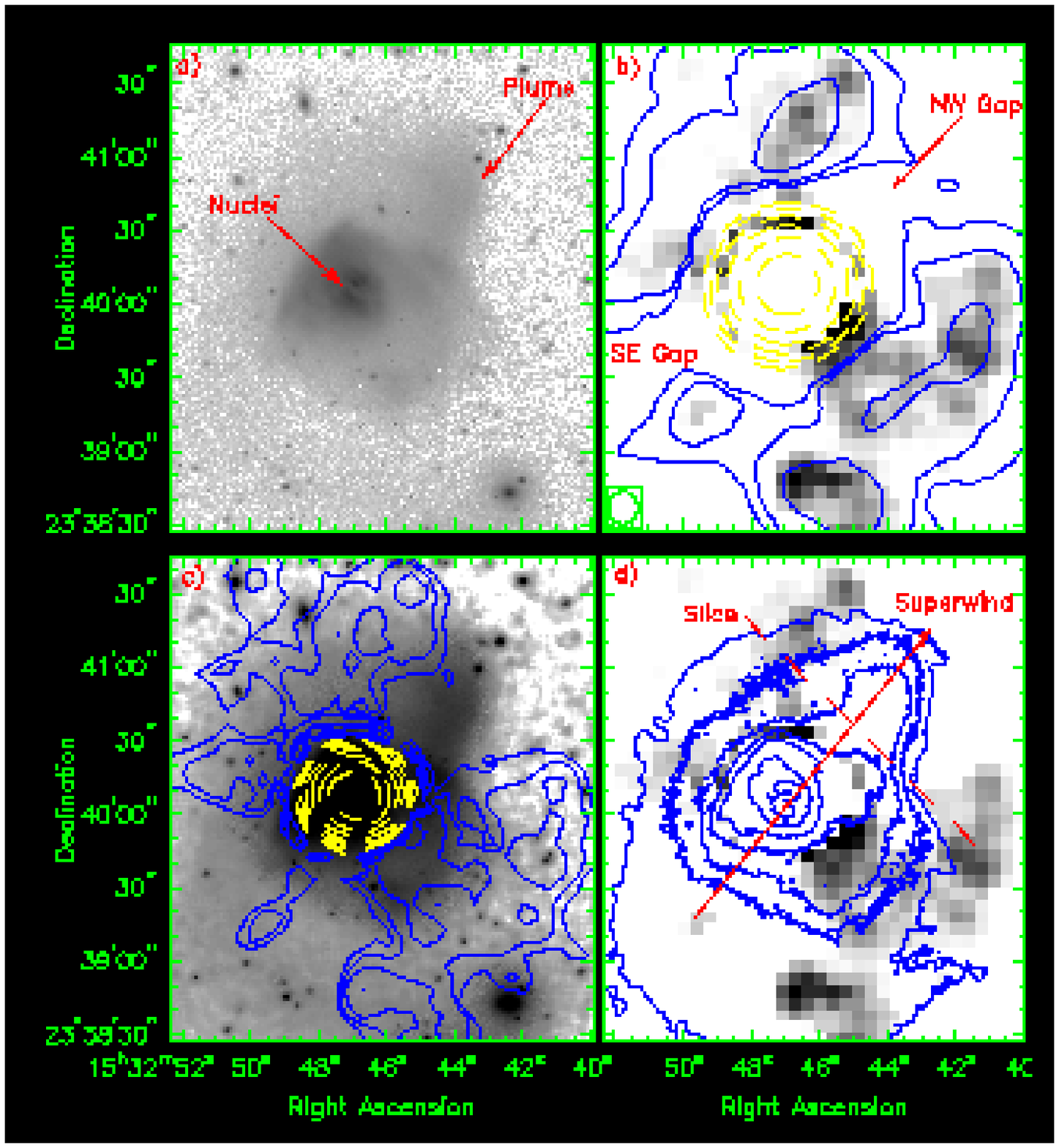}{16.5cm}{0}{75}{75}{-250}{-90}
\caption{{\sc HI and Optical Morphology of Arp 220.} \small
Panel (a) (top left)
displays an \R-band image with features of interest labeled. Panel (b)
(upper right) presents a greyscale and contour representation of the 
\hi\  data. The contours indicate the distribution of \hi\  mapped at 
low-resolution ($\theta_{FWHM}\sim 30^{\prime\prime}$), whereas the
greyscales show the \hi\  mapped at higher resolution ($\theta_{FWHM}
\sim 18^{\prime\prime}$). The lowest contour is drawn at a column 
density of 5\col{19}, with successive contours a factor of 2 higher.
The greyscales range from 1\col{20} (white) to 4\col{20} (black). 
Panel (c) (lower left) presents the optical image in greyscales 
with contours from the 
high-resolution \hi\  data superimposed. The lowest contour is drawn 
at a column density of 5\col{19}, with successive contours a factor 
of 2 higher.  Panel (d) (lower right) shows contours from a
star-subtracted \R-band image (lowest contour of 26 \msqas\  and contour
interval of 1 \msqas) upon greyscales of the high-resolution \hi\
data (from 1\col{20} to 4\col{20}). The dashed line labeled ``Slice"
indicates the locations of the intensity profiles plotted in
Fig.~\ref{fig:slices}c. The arrow labeled ``Superwind" indicates the 
direction (but not the extent) of the putative expanding superwind.}
\label{fig:a220mos}
\end{figure*}

Arp 220 (UGC~9913, IC~4453/4) is the prototypical ultraluminous
infrared galaxy with $L_{8-100\mu m}=1.5 \times 10^{12} L_\odot$
(Soifer et al.  1984).  It is an advanced merger system with two radio
and infrared nuclei separated by 0.9\as\  (345 pc for $D$= 79 Mpc), and
a bright optical plume extending 35 kpc to the NW
(Fig.~\ref{fig:a220mos}a).  Each of the two nuclei has its own compact
molecular disk. The two nuclear disks are in turn embedded in one
larger 1 kpc scale molecular gas disk (see Scoville, Yun, \& Bryant
1997, Downes \& Solomon 1998, Sakamoto \et\  1999 and references
therein).  The spin axis of the eastern nucleus is aligned with that
of the kpc-scale disk while the western nucleus rotates in the
opposite direction.  These observations suggest that Arp 220 is the
product of a prograde-retrograde merger of two gas rich spiral
galaxies (Scoville, Yun, \& Bryant 1997). 

An irregular disk-like distribution of neutral hydrogen extends over a
100 kpc diameter region surrounding the optical galaxy (Yun \& Hibbard
1999a). The overall \hi\  kinematics indicates that this material has a
component of rotation in the same sense as that for the eastern
nucleus and the molecular gas disk, and opposite the rotation of the
western nucleus.  This suggests that the \hi\  disk and eastern nucleus
originated from the retrograde progenitor, while the the western
nucleus and NW optical plume (Fig.~\ref{fig:a220mos}a) arose from the 
prograde progenitor.

Because of the vigorous star formation occurring within Arp 220
(SFR=340 \Myr, HAM90), much of the \hi\   within the optical body of the
system is seen only in absorption against the bright radio continuum
emission from the central starburst.  Beyond this, the \hi\   has high
column densities (\NHI\about 1.5\col{20}), but only to the NE and
SW. Most notably, there are local \hi\   minima to the NW and SE (see
gaps in Fig.~\ref{fig:a220mos}b). Comparison of the \hi\   map with the
optical image (Fig.~\ref{fig:a220mos}c) shows that the NW gap occurs
exactly at the location of the optical tail. The relationship between
the optical and \hi\   surface brightness levels across this feature are
illustrated by an intensity profile measured along the dotted line
shown in Fig.~\ref{fig:a220mos}d, and plotted in Fig.~\ref{fig:slices}c.  
As in NGC 520 and Arp 299, the gas column density increases precisely
where the optical light from the tail begins to fall off.  
There is a similar \hi\ gap to the SE, but in this case there is no
corresponding optical feature associated with it.  At even larger
radii, the \hi\ is more diffuse (\NHI\about 3\col{19}) and has no
optical counterpart down to \mR=27 \msqas.

An X-ray image obtained with the ROSAT HRI camera (Heckman \et\   1996)
reveals an extended central source that is elongated along P.A.=135\fdg\
(indicated by arrows in Fig.~\ref{fig:a220mos}).  A deep \HaN\   image
of Arp 220 reveals ionized gas with a bright linear morphology at this
same position angle (Heckman, Armus \& Miley 1987).  The optical
emission line kinematics are suggestive of a bipolar outflow (HAM90),
and the physical properties of the warm and hot gas strongly support
the superwind scenario for this emission (HAM90, Heckman \et\   1996).  
As in NGC 520 and Arp 299, the position angle of the putative expanding 
superwind is in the same direction as the \hi\ minima, i.e.~NW and SE.

\section{Discussion}
\label{sec:disc}

\begin{figure*}[t]
\plotfiddle{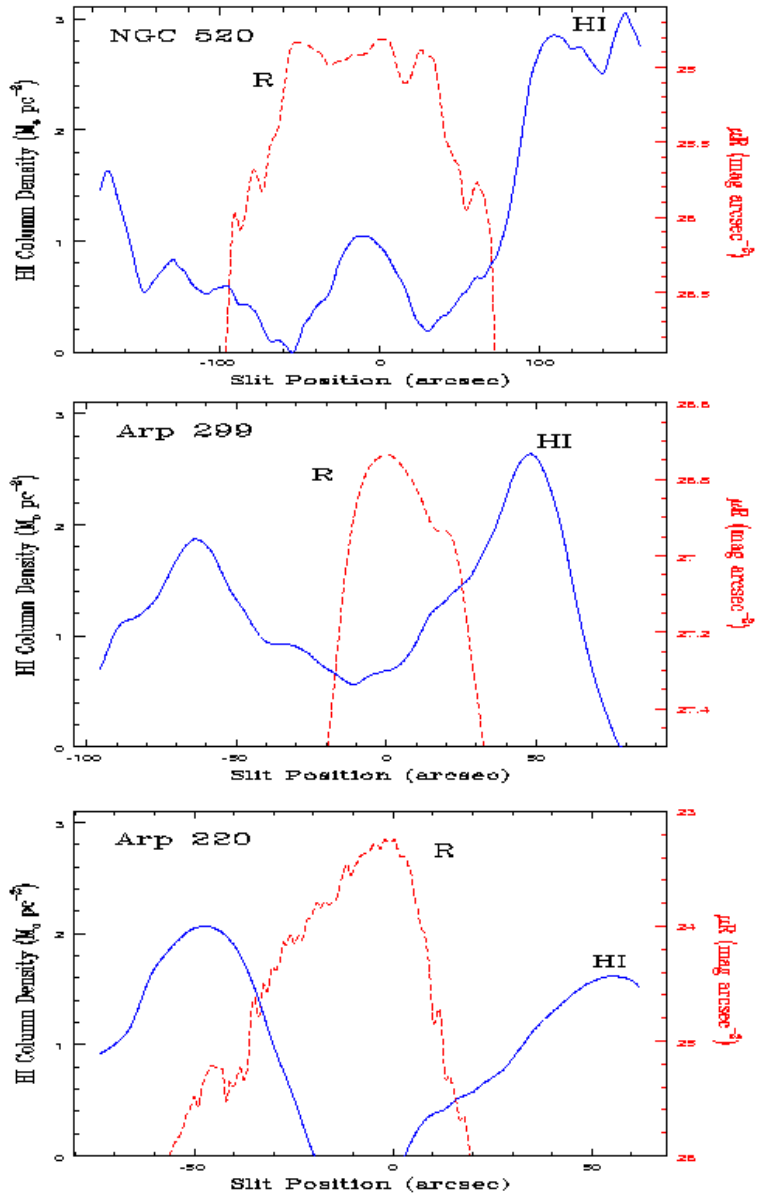}{16cm}{0}{65}{65}{-160}{10}
\caption{\small
Intensity profiles of \R-band surface brightness ($\mu_R$) and \hi\   
column density (beam-averaged values measured from the low resolution 
data) taken along the dashed lines in Figs.~\ref{fig:n520mos}--
\ref{fig:a220mos}.  (a)~~(top) NGC 520, across the northern plume.  
(b)~~(middle) Arp 299, across the tip of the northern tail.  (c)~~
(bottom) Arp 220, across the NW plume. Gas column densities are measured 
in $M_\odot \, {\rm pc}^{-2}$ (where 1 $M_\odot \,{\rm pc}^{-2}$ = 
1.25\col{20}), and optical surface brightnesses in mag arcsec$^{-2}$.}
\label{fig:slices}
\end{figure*}

Figures~\ref{fig:n520mos}--\ref{fig:slices} provide evidence for both
small- and large-scale differences in the distributions of the tidal
gas and stars in these three systems.  The small-scale differences are
of the type illustrated in Fig.~\ref{fig:slices}, whereby the gas
column density falls off just as the optical surface brightness
increases at various edges of the tidal features.  
In NGC 520 and Arp 220, the large-scale differences are between the
outer \hi\   rings and disks (which have no associated starlight) and the
optical tails and plumes (which have no associated \hi).  Although these
features are kinematically decoupled at present (with the gas rings and
disks predominantly in rotation and the optical tails and plumes
predominantly in expansion), it is possible that they had a common
origin and have subsequently decoupled and evolved separately.  In Arp
299, on the other hand, the \hi\   filaments and optical tail have similar
morphologies and continuous kinematics and are therefore part of the
same kinematic structure.  In this system we believe the bifurcated
tidal morphology results from a progenitor with a warped gaseous disk
(\S\ref{sec:a299} \& HY99), and we seek to understand why the inner
filament is gas-poor, given that its progenitor was obviously gas-rich.

In this section we investigate a number of possible explanations for
these observations. In particular, we discuss the possible role played
by: differences in the initial radial distribution of the gas and
stars (\S \ref{sec:radial}), dust obscuration (\S \ref{sec:dust}),
kinematic decoupling of the gas due to collisions within the
developing tidal tail (\S \ref{sec:coll}), ram pressure stripping of
the gas, either by a halo or by a galactic scale wind (\S
\ref{sec:RPS}), and photoionization of the gas, either by the
starburst or by local sources (\S \ref{sec:ioniz}).  

\subsection{Differences in the Radial Distribution of Gas and
Starlight}
\label{sec:radial}

In interacting systems the \hi\   is often more widely distributed than
the optical light (see, e.g. the \hi\   map of the M81 system by Yun
\et\   1994; see also van der Hulst 1979; Appleton, Davies \& Stephenson
1981). These gas-rich extensions frequently have no associated
starlight down to very faint limits (e.g. Simkin \et\   1986; HvG96).  A
natural explanation is that such features arise from the
\hi-rich but optically faint outer radii of the progenitor
disks. The relatively short lifetimes of luminous stars and the larger
velocity dispersions of less luminous stars, especially with respect
to the gas, will further dilute the luminous content of this material,
and the \hi -to-light ratio of the resulting tidal features will
increase with time (Hibbard \et\   1994).  Gaseous tidal extensions with
very little detectable starlight would seem to be the natural
consequence.  The outer \hi\   rings in NGC 520 and Arp 220 and the
gas-rich outer filament in Arp 299 are all likely to have arisen 
in this manner.

However, gas-rich outer disks cannot give rise to gas-poor optical
structures, such as the optical plume in NGC 520, the optical tail in
Arp 220, or the inner filament in Arp 299. Since these features
presumably arise from optically brighter regions of the progenitor
disks (regions which are characterized by \hi\  column densities higher
than that of the outer disks) one would have expected a priori that
these features should also be gas-rich. It is possible that the disks
which gave rise to the plumes in NGC 520 and Arp 220 were gas-poor at
all radii.  However, this would not account for the discontinuities in
the outer gaseous features that project near these optical features
(i.e., NE gap in NGC 520 and NW gap in Arp 220). We therefore seek
other explanations for these structures.

\subsection{Effects of Dust Obscuration}
\label{sec:dust}

The correspondence between rising gas column density and falling
optical surface brightness (Fig.~\ref{fig:slices}) suggests that dust
associated with the cold gas may attenuate the optical light.  To
address this possibility, we calculate the expected extinction in the
$R-$band for a given column density of \hi.  We adopt the Milky Way
dust-to-gas ratio determined by Bohlin, Savage, \& Drake (1978;
$N_{HI}/E(B-V)$= 4.8\col{21} mag$^{-1}$), which is supported by direct
imaging of the cold dust in the outer regions of eight disk galaxies
(Alton \et\   1998).  This is combined with the Galactic extinction law of
O'Donnell (1994; $A_R/E(B-V)=2.673$, from Table 6 of Schlegel \et\
1998) to yield an expected extinction in the \R-band of
$A_R={N_{HI}\over 1.8\times 10^{21} {\rm cm}^{-2}}$ mag.  

From Fig.~\ref{fig:slices}, the peak \hi\   column densities on either
side of the optical features are \about 3\col{20}.  The predicted
extinction is therefore of order 0.2 mag in the \R-band.  From
Fig.~\ref{fig:slices} we see that the mean light level drops by about
1.0 \msqas\   for Arp 299 (from 26.5 \msqas\   to below 27.5 \msqas),
about 1.5 \msqas\   for NGC 520 (from 25 \msqas\   to below 26.5 \msqas),
and by about 2.5 \msqas\   for Arp 220 (from 23.5 \msqas\   to below 26
\msqas) along the extracted slices. To produce this amount of extinction, 
the tidal gas would have to have a dust-to-gas ratio that is ten times 
that in the Milky Way.

The above analysis assumes that the measured neutral gas column density
represents the total gas column density.  However, the sharp drop in
\hi\   column density observed in many tidal features (HvG96, Hibbard 
\& Yun in preparation) suggests that the tidal gas may be highly ionized 
by the intergalactic UV field (see also references in \S \ref{sec:ioniz}).  
Since large dust grains should survive in the presence of this
ionizing radiation, the opacity per atom of neutral hydrogen
($A_R/N_{HI}$) should increase in regions of increasing ionization
fraction.  Observations of NGC 5018 (Hilker \& Kissler-Patig 1996), in
which blue globular clusters are absent in a region underlying an
associated \hi\ tidal stream, may support a high $A_R/N_{HI}$ ratio
for tidal gas.  Nevertheless, the lack of obvious reddening of the
$B-R$ colors along the slices in Arp 299 and NGC 520 (Hibbard 1995;
HY99) argues against a much higher extinction in these regions.

We conclude that extinction might be important for shaping the
morphology of the faintest optical features (e.g., the ``Hook'' and
the end of the optical tail of Arp 299, Fig.~\ref{fig:a299mos}, which
has \mR\   near the detection limit of 28 \msqas), but is insufficient
to greatly affect the overall tidal morphology.  However, an
anomalously high tidal dust-to-gas ratio remains a possibility.  This
question could be resolved by the direct detection of cold dust in
tidal tails with sub-millimeter imaging.

\subsection{Collisions within Developing Tidal Tails}
\label{sec:coll}

During the tail formation process, the leading-edge 
of the tail is decelerated with
respect to the center of mass of the progenitors, while the
trailing-edge is accelerated, and the two edges move towards each other
(see Toomre \& Toomre 1972, Fig.~3).  Eventually, the two edges appear
to cross, forming a caustic (Wallin 1990; Struck-Marcell 1990).  In most
cases, the caustics are simply due to projection effects.  Only for
low-inclination encounters will these crossings correspond to physical
density enhancements, and numerical experiments suggest that in these
cases the density will increase by factors of a few (Wallin 1990).  It
has been suggested that collisions experienced by the crossing tidal
streams in such low-inclination encounters may lead to a separation
between the dissipational (gas) and non-dissipational (stellar) tidal
components (Wevers 1984; Smith \et\   1997). 

The present data do not allow us to directly address this question,
since the kinematic decoupling presumably took place long ago.  However,
several arguments lead us to suspect that this collisional process is
not important in tidal tails: (1) large scale decoupling between the
stellar and gaseous tidal morphologies is not seen in many systems known
to have experienced low-inclination encounters (e.g.  NGC 4038/9, ``The
Antennae'', Hibbard \et\   in preparation; NGC 7252 ``Atoms for Peace''
Hibbard \et\   1994; NGC 4676 ``The Mice'' HvG96); (2) the broad
plume-like morphologies of the optical features in Arp 220 and NGC 520
suggest rather inclined encounters (\S\S \ref{sec:n520}, \ref{sec:a220}),
in which case wide-spread collisions are not expected; and (3) the
parallel filaments in the Arp 299 tail have identical kinematics,
whereas one would expect kinematic differences between the stripped and
unstripped material. 

Therefore while gaseous collisions and dissipation might result in
differences between gas and stars during tidal development
(particularly along tidal bridges, where the gas streamlines are
converging; e.g. Struck 1997; NGC 7714/5 Smith \et\   1997; Arp 295
HvG96), we believe that they are not likely to lead to a wide-spread
decoupling in the outer regions.

\subsection{Ram Pressure Stripping}
\label{sec:RPS}

If the tidal features pass through a diffuse warm or hot medium, or if
such a medium passes through the tidal features, it is possible that
the tidal gas exchanges energy and momentum with this medium
due to collisions.  Such effects have been proposed to explain the
stripping of the cool interstellar medium from spiral galaxies as they
move through the hot IGM in clusters (Gunn \& Gott 1972), and is
referred to as Ram Pressure Stripping (RPS).  Tidal features should 
be relatively easily stripped, as they lack the natural restoring
forces present in disk galaxies, except possibly at a small number of
self-gravitating regions.  In this case, the momentum imparted due to
ram pressure is simply added to or subtracted from the momentum of the
gaseous tidal features, and a separation of stellar and gaseous
components might be expected.

In the next two subsections, we investigate two possible sources for ram
pressure: an extended halo associated with the progenitors (\S
\ref{sec:RPShalo}); and an expanding starburst driven superwind (\S
\ref{sec:RPSwind}). 

\subsubsection{RPS from Extended Halo Gas}
\label{sec:RPShalo}

Our own galaxy is known to have an extended halo of hot gas (Pietz \et\
1998).  The existence of similar halos around external galaxies has been
inferred from observations of absorption line systems around bright
galaxies (e.g.~Lanzetta \et\   1995).  These halos may have sufficient
density to strip any low column density gas moving through them. 
Several investigators have suggested that such stripping is responsible
for removing gas from the Magellanic Clouds as they orbit through the
Galaxy's halo, producing the purely gaseous Magellanic Stream
(e.g.~Meurer, Bicknell \& Gingold 1985; Sofue 1994; Moore \& Davis
1994).  Sofue \& Wakamatsu (1993) and Sofue (1994) specifically stress
that stripping by galaxy halos should also play an important role in the
evolution of \hi\   tidal tails. 

The tidal features in each of our systems have \hi\   column densities
and velocities similar to those assumed in the numerical models of
Sofue (1994) and Moore \& Davis (1994), which resulted in rather
extreme stripping of the \hi\   clouds. Although this may seem to
provide an immediate explanation for our observations of gas/star
displacements, we point out that these column densities and velocities
are typical of all of the tails thus far imaged in \hi, the great
majority of which do not show the extreme displacements we describe
here.  There is no reason to believe that the halo properties of NGC
520, Arp 299 and Arp 220 are any different from, or that the
encounters were any more violent than similar mergers which do not
exhibit such dramatic displacements (e.g., NGC 4038/9, NGC 7252, NGC
4676 HvG96; NGC 3628 Dahlem \et\   1996; NGC 2623, NGC 1614, Mrk 273
Hibbard \& Yun in preparation; NGC 3256 English \et\   1999). 
In fact,  in light of the stripping simulations mentioned above, 
one wonders why such displacements are not more common. 

A possible solution to this puzzle is suggested by the results of
numerical simulations of major mergers.  In these simulations, the
material distributed throughout the halos of the progenitors is
tidally distended along with the tails, forming a broad sheath around
them (see e.g.~the video accompanying Barnes 1992). This sheath has
similar kinematics as the colder tail material, resulting in much
lower relative velocities than if the tail was moving through a static
halo, thereby greatly reducing any relative ram pressure force.

In summary, while halo stripping might be effective for discrete
systems moving through a static halo (such as the LMC/SMC through the
halo of the Galaxy, or disks through a hot cluster IGM), the lack of
widespread \hi/optical decoupling in mergers suggests that it is not
very effective for removing gas from tidal tails, and it does not
appear to be a suitable explanation of the present observations.

\subsubsection{RPS from Expanding Superwind}
\label{sec:RPSwind}

The three systems under discussion host massive nuclear starbursts with
associated powerful outflows or ``superwinds".  Optical emission lines
and/or X-ray emission reveal that the observed outflows extend for tens
of kpc from the nuclear regions.  Theoretical calculations suggest that
the observed gas plumes represent just the hottest, densest regions of a
much more extensive, lower density medium (Wang 1995).  In each of these
three systems, the most extreme gaps in the \hi\   distribution appear
along the inferred direction of the expanding hot superwind
(Figs.~\ref{fig:n520mos}d, \ref{fig:a299mos}d, \ref{fig:a220mos}d).  A
very similar anti-correlation between an out-flowing wind and tidally
disrupted \hi\  has been observed in the M82 system (Yun \et\   1993;
Strickland \et\   1997; Shopbell \& Bland-Hawthorn 1998), the NGC 4631
system (Weliachew \et\   1978; Donahue \et\   1995; Wang \et\   1995; Vogler
\& Pietsch 1996), and possibly the NGC 3073/9 system (Irwin \et\   1987;
Filippenko \& Sargent 1992).  It has been suggested that this
anti-correlation is due to an interaction between the blown-out gas of
the superwind and the cold gaseous tidal debris, either as the wind
expands outward into the debris, or as the tidal debris passes through
the wind (Chevalier \& Clegg 1985; Heckman, Armus \& Miley 1987, and
references above). 

\begin{figure*}[t]
\plotfiddle{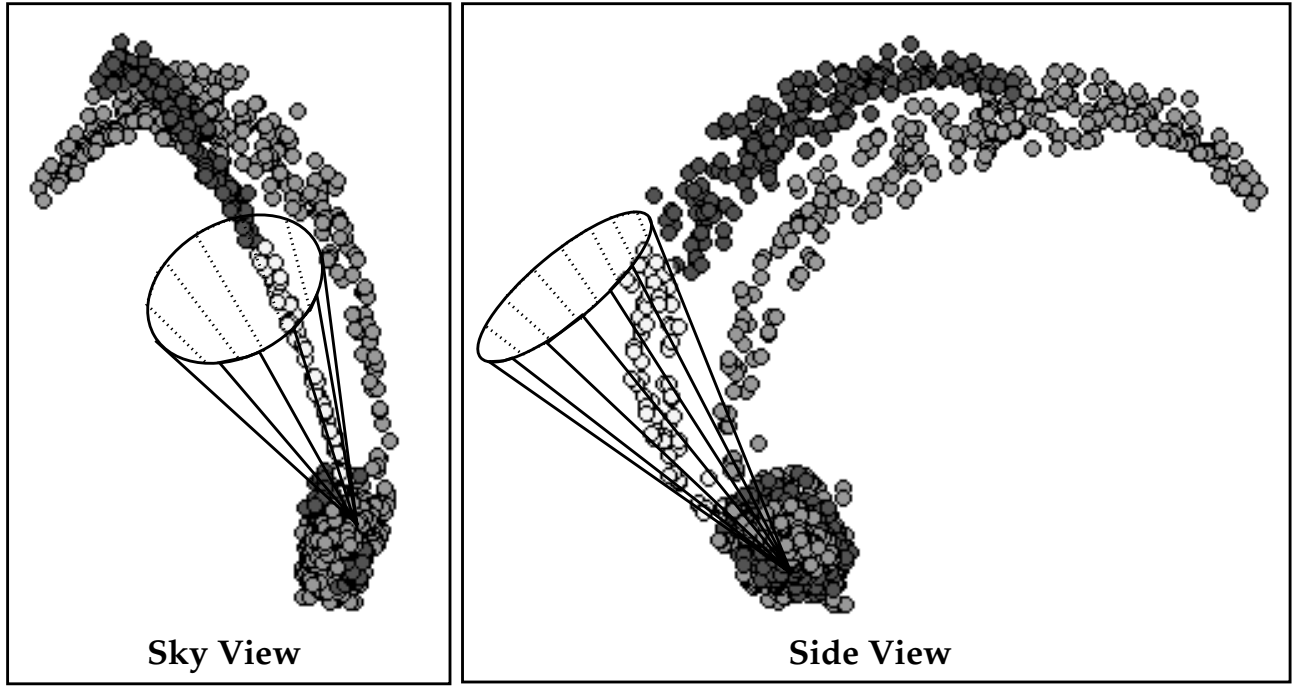}{10cm}{0}{60}{60}{-240}{20}
\caption{\small
An illustration of the proposed relative placement of the tidal tail 
and the expanding wind or ionization cone for the case of Arp 299. 
In this figure, the purely gaseous tidal filament is represented by 
light grey circles, whereas the optical tidal filament is represented 
by circles drawn with two shades: regions of the optical tail which have 
accompanying \hi\ are shaded dark grey, while the regions which 
have been cleared of \hi\ are shaded white (see Fig.~\ref{fig:a299mos}). 
The suggested geometry of the superwind bubble (cf. \S \ref{sec:RPSwind})
or ionization cone (cf. \S \ref{sec:ionizSB}) is indicated by the
cone, which has been drawn to encompass the white circles in the optical
filament. This figure illustrates how the restricted opening angle of 
such a cone may only intersect a portion of the ribbon-like 
tail. }
\label{fig:cartoon}
\end{figure*}

Figure \ref{fig:cartoon} presents the suggested geometry for the
case of Arp 299. This figure is constructed from our preliminary 
efforts to model the northern tail of Arp 299 using N-body 
simulations similar to those presented in Hibbard \& Mihos (1995; 
i.e. no hydrodynamical effects are included). We found that we 
could not match the morphology and kinematics of both filaments 
simultaneously, but could match either one separately.
Fig.~\ref{fig:cartoon} presents the results of combining these
two solutions. In this sense this figure is {\it not a self-consistent 
fit to the data}, but simply a cartoon which illustrates the proposed 
relative placement of the tidal tail and the expanding wind. In this 
figure, the wind opening angle is illustrated by the cone, the 
gas-rich regions of the tails are represented by dark and light 
grey circles, and the gas-poor regions of the tails are represented 
by white circles.  The figure illustrates how the restricted
opening angle of such a wind (or of an ionization cone, cf. \S
\ref{sec:ionizSB}) may intersect only a portion of the ribbon-like
tail.

Here we estimate whether ram pressure stripping by the nuclear
superwind can exert sufficient pressure on gas at the large distances
typical of tidal tails.  We use equation (5) from Heckman, Lehnert \&
Armus (1993; see also Chevalier \& Clegg 1985) to calculate the
expected ram pressure ($P_{RPS}$) of the superwind far from the
starburst as a function of its bolometric luminosity ($L_{bol}$):

\begin{equation}
P_{RPS}(r) = 
4\times 10^{-10} \,{\rm dyne \, cm}^{-2} 
\left({L_{bol}\over 10^{11} L_\odot}\right)
\left({1 \,{\rm kpc}\over r}\right)^2
\label{eq:Prps}
\end{equation}

This equation has been shown to fit the pressure profile derived from
X-ray and optical emission line data of Arp 299 (Heckman \et\   1999).  
It should provide a lower limit to the ram pressure, since it assumes 
that the wind expands spherically, while observations suggest that the 
winds are limited in solid angle. 

This pressure can be compared with the pressure of the ambient medium in
the tidal tail ($P_{tidal}$), given by the energy per unit volume: 

\begin{equation}
P_{tidal} = C \times \rho_{gas} \sigma_{gas}^2
\label{eq:perfgas}
\end{equation}

\noindent where $C$ is a constant, $\rho_{gas}$ is the mass density of
the gas, and $\sigma_{gas}$ is the velocity dispersion of the gas.
For an equation of state of the form $P\sim\rho^\gamma$, the constant
$C$ is equal to $\gamma^{-1}={3\over 5}$, and $\sigma_{gas}$
corresponds to gas sound speed, while for a self-gravitating cloud, $C={3
\over 2}$ and $\sigma_{gas}$ is the one-dimensional velocity dispersion
of the cloud. In both cases, we assume that the observed line-of-sight 
velocity dispersion of the \hi\   is a suitable measure of $\sigma_{gas}$. 

The mass density of the tidal gas is given by $\rho_{gas}=1.36 \times
m_H \times n_{HI}$, where the numerical constant accounts for the
presence of {\smcap H}e, $m_H$ is the mass of a hydrogen atom, and
$n_{HI}$ is the number density of atomic hydrogen.  If we assume the
gas is uniformly distributed\footnote{Clearly the results will be very
different if the tidal gas is mainly in dense clouds, a point that can
be tested with higher-resolution VLA observations. For now we
calculate the ram pressure effect on the diffuse gas.} with a column
density $N_{HI}$ along a length $dL$, we have $n_{HI} = 6.5\times
10^{-3} \, {\rm cm}^{-3} ({N_{HI} \over 2\times 10^{20} \, {\rm
cm}^{-2}}) ({10 \, {\rm kpc} \over dL})$, where the fiducial values
are typical of tidal features (HvG96, Hibbard \& Yun in preparation).
We rewrite eqn.~(\ref{eq:perfgas}) in terms of the observables:

\begin{eqnarray}
\lefteqn{ {P_{tidal} } =  
2\times 10^{-14} \,{\rm dyne \, cm}^{-2} \,
\left({C\over 1.5}\right) } \nonumber\\
& & 
\left({N_{HI}\over 2\times10^{20} {\rm cm}^{-2}}\right)
\left({10 \,{\rm kpc} \over dL}\right)
\left({\sigma_{HI} \over 10 \,{\rm km \, s}^{-1}}\right)^{2}
\label{eq:Ptidal}
\end{eqnarray}

The maximum radius out to which we expect material to be stripped
($R_{RPS}$) is then given by the requirement that $P_{RPS}$(r)
$=P_{tidal}$(r) at r$=R_{RPS}$.  We replace $L_{bol}$ in
eqn.~(\ref{eq:Prps}) by the IR luminosity ($L_{IR}$), under the
assumption that the IR luminosity arises from reprocessed UV photons
from the starburst\footnote{This assumes that the IR luminosity is not 
enhanced due to the presence of AGN. There is no evidence for an 
energetically important AGN in any of these three systems.}  
(Lonsdale, Persson \& Matthews 1984, Joseph \&
Wright 1985).  The very high IR luminosities of these systems
($L_{IR}/L_B>10$) make it likely that this is indeed the case, and we
are probably making an error of $\lesssim$ 10\% (e.g. Heckman, Lehnert
\& Armus 1993). Equating eqns.~(\ref{eq:Prps}) and
(\ref{eq:Ptidal}) we find:

\begin{eqnarray}
\lefteqn{ {R_{RPS} } = 
140 \,{\rm kpc} \,
\left({1.5\over C}\right)^{1 \over 2}
\left({L_{IR}\over 10^{11} L_\odot}\right)^{1 \over 2}} \nonumber\\
& & 
\left({2\times10^{20} {\rm cm}^{-2} \over N_{HI}}\right)^{1 \over 2}
\left({dL \over 10 \,{\rm kpc}}\right)^{1 \over 2}
\left({10 \,{\rm km \, s}^{-1} \over \sigma_{HI}}\right)
\label{eq:Rrps}
\end{eqnarray}

In Table~\ref{tab:results} we provide estimates of $R_{RPS}$ for the
systems considered here. In calculating $R_{RPS}$ we made the very
conservative assumption that the values of $N_{HI}$ and $\sigma_{HI}$
for the stripped gas are equal to the maximum values found within the
tidal tails (see Table~\ref{tab:results}).  The results of these
calculations indicate that, in all cases, $R_{RPS}$ is larger than the
radii of the observed gaps in the tidal \hi\ distributions. Therefore,
in principle, the wind should be able to strip the gas from any tidal
material in its path.

The above derivation assumes that the tidal gas is at rest with respect
to the wind.  It can be easily generalized to the case of a wind
impacting an expanding tidal feature by reducing the wind ram pressure
by a factor $({V_{wind}-V_{tidal} \over V_{wind}})^2$.  For NGC 520 and
Arp 220, the tidal gas is primarily in rotation (\ie, moves
perpendicular to superwind), so we expect the gas to feel the full ram
pressure given above.  For Arp 299, Heckman \et\   (1999) find $V_{wind}$
= 800 \kps, and we estimate a maximum $V_{tidal}$ = 240 \kps\   (HY99). 
Therefore, the ram pressure could be reduced by 50\%, reducing $R_{RPS}$
by 70\% from that listed in Table~\ref{tab:results}, \ie\   $R_{RPS} \sim
100$ kpc for Arp 299.  This is still large enough to reach to the region
of the N clump in Fig.~\ref{fig:a299mos}.  The regions of high \hi\
velocity dispersion indicated in Fig.~\ref{fig:a299mos}c may be due to
the influence of such a wind.  We note that these regions occur on the
side of the \hi\   features that face the starburst region.  However, no
such kinematic signatures are visible in the gas near the wind axis in
NGC 520 or Arp 220. 

The lack of gaseous/stellar displacements in the tidal tails of many
superwind systems might seem to provide a strong argument against the
scenario outlined above.  However, there are two conditions needed to
produce wind-displaced tidal features: the starburst must be of
sufficient energy and duration to achieve ``blowout", and the tidal 
\hi\  must intersect the path of the expanding wind material. The 
second condition is not met for the blowout systems NGC 4676, NGC
3628, NGC 2623, NGC 1614, and NGC 3256 (references given in \S
\ref{sec:RPShalo}).  In these systems the tidal tails appear to lie at
large angles with respect to the blowout axis, and their tidal tails
should not intersect the wind.  Both conditions are met for M82 and
NGC 4631, which both show extreme \hi/optical displacements.  For
these two systems, the high-latitude \hi\ appears to be accreted from
nearby disturbed companions (M81 and NGC 4656, respectively; Yun \et\
1993, Weliachew \et\ 1978), while for the three major mergers under
study here the \hi\ appears to intercept the path of the wind as a
result of a highly inclined encounter geometry. A high inclination
encounter geometry is therefore a prerequisite for such displaced
morphologies, and the host of the superwind should be the disk with a
retrograde or polar spin geometry.

Nevertheless, ram pressure stripping cannot provide a complete
explanation of the observations.  Since the stars are unaffected by
the wind, it would be an unusual coincidence for the edges of the
optical plumes to correspond with the edge of the cold gas which is
presently being ablated.  Nor does it seem likely that the wind could
be sufficiently collimated to ``bore" into the northern \hi\   clump in Arp
299 just where the optical tail appears projected upon it.  Therefore
a second process is still needed to explain the small-scale
anti-correlations.  In conclusion, the expanding wind {\it should}
affect any tidal \hi\   in its path; however this effect
alone cannot explain all the details of the observations.

\subsection{Photoionization}
\label{sec:ioniz}

Disk galaxies are known to exhibit a precipitous drop in neutral
hydrogen column density beyond column densities of a few times $10^{19}$
cm$^{-2}$ (Corbelli, Schneider \& Salpeter 1989; van Gorkom 1993;
Hoffman \et\ 1993).  This drop has been attributed to a rapid change
in the ionization fraction of the gas due to influence of the
intergalactic UV field (Maloney 1993, Corbelli \& Salpeter 1993, Dove
\& Shull 1994; see also Felten \& Bergeron 1969, Hill 1974), rather
than a change in the total column density of H.

Tidal tails are assembled from the outermost regions of disk galaxies.
Since this material is redistributed over a much larger area than it
formerly occupied, its surface brightness must decrease accordingly.
Therefore, if the progenitors were typical spirals, with \hi\ disks
extending to column densities of a few times $10^{19}$ cm$^{-2}$, then
the resulting tidal tail {\it must} have gas at much lower column
densities.  However, tidal tails exhibit a similar edge in their
column density distribution {\it at a similar column density} (HvG96;
Hibbard \& Yun in preparation).  This is one of the most compelling
pieces of evidence for an abrupt change in the phase of the gas at low
\hi\ column density.  The outer tails mapped in \hi\ should therefore
be the proverbial ``tip of the iceberg'' of a lower column density,
mostly ionized medium.  With the tidal gas in this very diffuse state,
fluctuations in the incident ionizing flux might be expected to
produce accompanying fluctuations in the neutral gas fraction.

Given these considerations, we examine the possibility that the total
hydrogen column density does not change at the regions illustrated in
Figs.~\ref{fig:n520mos}--\ref{fig:slices}, but that the neutral fraction
does, \ie\   that the gas in the regions under study has a higher
ionization fraction than adjacent regions.  The intergalactic UV field
should be isotropic, and would not selectively ionize certain regions of
the tails.  Here we examine the possibility of two non-isotropic sources
of ionizing flux: (1) leakage of UV flux from the circumnuclear
starburst; (2) ionization by late B stars and white dwarfs
associated with the evolved stellar tidal population. 

Our procedure is to compare the expected ionizing flux density
shortward of 912 \AA\  to the expected surface recombination rate of
the gas. We assume that in the area of interest the gas is at a
temperature of $\sim 10^4$ K, for which a case-B recombination
coefficient of $\alpha_B = 2.6\times 10^{-13} \, {\rm cm^3~s}^{-1}$ is
appropriate (Spitzer 1956). We assume that the hydrogen is almost
completely ionized, so $n_e \approx n_H$. We further assume that the
density of ionized gas is the same as the density of the neutral gas
in the adjacent regions, $n_H \approx n_{HI}$, where $n_{HI}$ is
calculated as above ($n_H \sim N_{HI} / dL$, \S \ref{sec:RPSwind}). 
The detailed ionization state will depend sensitively on the
clumpiness of the gas, but a full treatment of this problem is beyond
the scope of this paper. Here we wish to investigate if these
processes are in principle able to create effects similar to those
observed.

\subsubsection{Photoionization by the Starburst}
\label{sec:ionizSB}

Here we consider the case that the superwind does not affect the tidal
gas by a direct interaction, but influences it by providing a direct
path from the tidal regions to the starburst, free of dust and dense
gas (see also Fig.~\ref{fig:cartoon}).
Through these holes, ionizing photons from the young hot stars
stream out of the nuclear regions and are quickly absorbed by the
first neutral atoms they encounter. Following Felton \& Bergeron (1969;
see also Mahoney 1993), we solve the equation:

\begin{equation}
n_{HI}^2 \alpha_B dL = I = {1\over 4 \pi
r^2} \int^\infty\limits_{h\nu=13.6 eV} {L_\nu \over h\nu} d\nu
\label{eq:photoSB}
\end{equation}

The right hand side of this equation represents the total ionizing
radiation escaping the starburst region along a direction that has been
cleared of obscuring material by the superwind.  We express this in
terms of the total ionizing flux of a completely unobscured starburst of
a given bolometric luminosity $L_{bol}$ by introducing the factor
$f_{esc}(\Omega_{wind})$ to account for the fact that only a fraction of
the photons emitted into a solid angle $\Omega_{wind}$ find their way
out of the starburst region\footnote{It is important to differentiate
$f_{esc}$, the total fraction of ionized photons emerging from a
starburst, and $f_{esc}(\Omega_{wind})$, the fraction emerging along a
particular sightline.  $f_{esc}$ is the total angle averaged fraction, i.e. 
$f_{esc}$ is the integral of $f_{esc}(d\Omega)$ over all solid angles,
while $f_{esc}(\Omega_{wind})$ is the integral over a solid angle
cleared by the wind. Most studies in the literature quote values for 
$f_{esc}$.}. The expected ionizing flux for a starburst of a
given bolometric luminosity $L_{bol}$ is calculated from the population
synthesis models of Bruzual \& Charlot (1993; 1995), assuming continuous
star formation with a duration longer than 10 Myr (long enough for the
burst to achieve blowout), a Salpeter IMF with $M_{lower}=0.1 M_\odot$
and $M_{upper}=125 M_\odot$, and solar metallicity.  This
yields\footnote{With similar assumptions, the ``Starburst99'' models of
Leitherer et al.\   (1999) give approximately the same numerical
coefficient.} $I = f_{esc}(\Omega_{wind}) \times {1.83\times 10^{54}
\,{\rm photons \, s}^{-1} \over 4 \pi r^2} \times {L_{bol}\over 10^{11}
L_\odot}$.  Again making the standard assumption that most of the
starburst luminosity is emitted in the far infrared (\ie\
$L_{bol}=L_{IR}$, cf.  \S \ref{sec:RPSwind}), we rearrange
eqn.~(\ref{eq:photoSB}) to solve for the radius, $R_{ionized}$, out to
which the starburst is expected to ionize a given column density of \hi\
of thickness $dL$:

\begin{eqnarray}
R_{ionized} & = & 66 \,{\rm kpc} \,
\left({f_{esc}(\Omega_{wind}) \over 0.10}\right)^{1/2}
\left({L_{IR}\over 10^{11} L_\odot}\right)^{1/2}\nonumber\\
& & \left({2\times10^{20} {\rm cm}^{-2} \over N_{HI}}\right) 
\left({dL \over 10 \,{\rm kpc}}\right)^{1/2}
\label{eq:Rion}
\end{eqnarray}

Resulting values for $R_{ionized}$ are listed in
Table~\ref{tab:results}.  For this computation, we have adopted a value
of 10\% for $f_{esc}(\Omega_{wind})$.  This is equal to the total
fraction of ionizing photons, $f_{esc}$, escaping from a normal disk
galaxy as calculated by Dove, Shull \& Ferrara (1999).  Even higher
values of $f_{esc}$ are expected in starburst systems (Dove \et\   1999). 
Since we stipulate that a higher fraction of ionizing photons escape
along sightlines above the blowout regions than are emitted along other
directions, it follows that $f_{esc}(\Omega_{wind}) \, > \, f_{esc}$,
and as a result the values of $R_{ionized}$ calculated in
Table~\ref{tab:results} should be conservative estimates. 

Table~\ref{tab:results} shows that under these simplified conditions,
$R_{ionized}$ is of the order of, or larger than, the tidal radii of
interest.  We therefore conclude that the starburst seems quite capable
of ionizing tidal \hi, if indeed there is an unobstructed path from the
starburst to the tidal regions.  This might explain the lack of \hi\
along the wind axis in NGC 520 and Arp 220, and the absence of \hi\
along the optical tidal tail in Arp 299.  

This process is especially attractive since it can potentially
explain the lack of \hi\   at the bases of otherwise gas-rich tidal 
tails in NGC 7252 (Hibbard \et\ 1994), Arp 105 (Duc \et\   1997), and 
NGC 4039 (Hibbard, van der Hulst \& Barnes in preparation). 
These systems do {\it not} show evidence for expanding superwinds,
which rules out the possibility that RPS is playing a role. And each of 
these systems possesses a level of star formation that, according 
to eqn.~\ref{eq:Rion}, is capable of ionizing gas out to the necessary
radii. 

However, photoionization by the central starburst does not seem
capable of explaining all of the observations.  As with the wind
hypothesis above, it would be an unusual coincidence for the edges of
the optical plumes to correspond with the edge of ionization cone.  
Therefore a second process is still needed to explain the
small-scale anti-correlations. 

\subsubsection{Photoionization by the Optical Tails}
\label{sec:ionizTAILS}

The fact that the \hi\ column density falls off just as the optical
surface brightness increases at the edges of various tidal features
(Fig.~\ref{fig:slices}) leads us to suspect that there may be local
sources of ionization within the stellar features themselves. For NGC
520 and Arp 220, we believe the outer \hi\ is in a disk structure
which is intersected by the tidal stellar plumes and we wish to
investigate whether ionization by evolved sources within the stellar
plumes, such as late B stars and white dwarfs, could be responsible
for decreasing the neutral fraction of the diffuse outer \hi. For Arp
299 the geometry is more complicated, and we refer the reader to
Fig.~\ref{fig:cartoon}. Here we suggest that part of the purely
gaseous tidal filament (the light grey filament in
Fig.~\ref{fig:cartoon}) is ionized by evolved sources near the end of
the stellar tail (the dark grey filament in Fig.~\ref{fig:cartoon},
especially those regions nearest the gas-rich filament in the right
hand panel of this figure).

As in the previous section, we balance the surface recombination rate
with the expected ionizing flux density (eqn.~\ref{eq:photoSB}).  In
this case, we calculate the ionizing flux density for an evolved
population of stars of a given \R-band luminosity density ($\Sigma_R$,
in $L_\odot \,{\rm pc}^{-2}$).

In order to approximate the stellar populations in the tidal tails, we
assume that the tails arise from the outer edges of an Sbc progenitor,
and that star formation ceased shortly after the tails were launched.
We again use the models of Bruzual \& Charlot (1993, 1995) for a Salpeter
IMF over the mass range 0.1--125 $M_\odot$, and adopt an exponentially
decreasing SFR with a time constant of 4 Gyr (typical of an Sbc
galaxy, Bruzual \& Charlot 1993), which is truncated after 10 Gyr and
allowed to age another 500 Myr. This simulates the situation in which
star formation within the disk is extinguished as the tail forms, and
the ejected stellar population passively fades thereafter. While tidal
tails frequently exhibit {\it in-situ} star formation (e.g. Schweizer
1978; Mirabel, Lutz \& Maza 1991), it is usually not widespread.
Under these assumptions, a population with a projected \R-band surface
brightness of 1 $L_{R,\odot}$ pc$^{-2}$ should produce an ionizing
flux of $2.36\times 10^4~ {\rm ph \,\, s^{-1} \, cm^{-2}}$. Therefore
eqn.~(\ref{eq:photoSB}) becomes $n_{HI}^2 \alpha_B dL < 2.36\times 10^4
{\rm ph \, s^{-1}\, cm}^{-2} \times \Sigma_R$, which can be rewritten
as:

\begin{eqnarray}
\Sigma_R 
& > &
14.28 \, L_{R,\odot} \,{\rm pc}^{-2} \,\,
\left({N_{HI}\over 2\times 10^{20} \,{\rm cm}^{-2}}\right)^2 
\left({10 \,{\rm kpc}\over dL}\right) 
\label{eq:SigmaR}
\end{eqnarray}

Noting that 1 $L_\odot \,{\rm pc}^{-2}$ corresponds to
\mR=25.9 \msqas, we rewrite this as a condition on the surface
brightness of the tidal features:

\begin{eqnarray}
\mu_R & < & 23.0 \, {\rm mag \, arcsec}^{-2} - \nonumber\\
& & 
2.5\times log \left[
\left({N_{HI}\over 2\times 10^{20} \,{\rm cm}^{-2}}\right)^2 
\left({10 \,{\rm kpc}\over dL}\right) \right]
\label{eq:muR}
\end{eqnarray}

Referring to Fig.~\ref{fig:slices}, we see that only the northern
plume of Arp 220 is bright enough to ionize nearby tidal \hi\
at the appropriate column densities. Neither the optical plume in the
NGC 520 system nor the northern tail in the Arp 299 system appears
bright enough to ionize the necessary columns of hydrogen unless the
tidal features are unreasonably thick (\about 60 kpc). However, 
since we have no other explanation for the small scale \hi/optical
differences illustrated in Fig.~\ref{fig:slices}, we are hesitant
to abandon this explanation too quickly. 

A possible solution is to invoke continued star formation even after
the tails are ejected.  For instance, if we do not truncate the star
formation rate after 10 Gyr, instead allowing the star formation rate
to continue its exponential decline as the tail expands, then the
ionizing flux per 1 $L_{R,\odot}$ pc$^{-2}$ is 70 times higher than
the value of $2.36\times 10^4~ {\rm ph \,\, s^{-1} \, cm^{-2}}$ used
above.  This would lower the fiducial surface brightness in
eqn.~(\ref{eq:muR}) from 23.0 \msqas\   to 27.5 \msqas, in which case
the faint tidal features in NGC 520 ($\mu_R \sim$ 25 \msqas) and Arp
299 ($\mu_R \sim$ 26.5 \msqas) could indeed ionize the necessary
column densities of adjoining \hi.

The observed broad-band colors of the tidal tails are not of
sufficient quality to discriminate between these two star formation
histories, since the expected color differences are only of order
$B-R$=0.1 mag. However, whether or not the gas is more highly ionized
in the regions of interest can be addressed observationally. The
expected emission measure ($EM=\int n_e^2 dl$) can be parameterized
as:

\begin{equation}
EM = 0.42 \, {\rm cm^{-6} pc} 
\left({N_{HI} \over 2\times 10^{20} \, {\rm cm}^{-2}}\right)^2 
\left({10 \, {\rm kpc} \over dL}\right)^2
\label{eq:EM}
\end{equation}

Since emission measures of order 0.2 cm$^{-6}$ pc have been detected
with modern CCD detectors (e.g.  Donahue \et\   1995; Hoopes, Walterbos \&
Rand 1999), there is some hope of being able to observationally
determine if regions of the tidal tails are significantly ionized.  If
the gas has a clumpy distribution, then there should be some high
density peaks which might be sufficiently bright to yield reliable
emission line ratios.  Such ratios would allow one to determine the
nature of the ionizing source, e.g.  photoionization vs.  shocks. 
Therefore, while we cannot assert unequivocally that photoionization
plays a role in shaping the outer tidal morphologies, it is possible to
test this hypothesis with future observations. 

The hypothesis that ionizing flux from a stellar tidal feature may 
ionize gas in a nearby gaseous tidal feature is not necessarily at odds 
with the observations that many stellar tails are gas-rich. This is 
because tails with cospatial gas and stars arise from regions originally
located within the stellar disk of the progenitors, while the optical
faint gas-rich tidal features likely arise from regions beyond the
optical disk (\S~\ref{sec:radial}). In normal disk galaxies, the \hi\
within the optical disk is dominated by a cooler component with a
smaller scale height and velocity dispersion, while the \hi\ beyond
the optical disk is warmer and more diffuse, with a larger scale
height and velocity dispersion (Braun 1995, 1997). As a result, $dL$
should be considerably larger for purely gaseous tidal features than 
for optically bright tidal features. 

\section{Conclusions}
\label{sec:concl}

In this paper we have described differences between gaseous and stellar
tidal features.  There are large-scale differences, such as extensive
purely gaseous tidal features (the outer disks in NGC 520 and Arp 220
and the outer filament in Arp 299) and largely gas-poor optical features
(tidal plumes in NGC 520 and Arp 220 and the inner filament in Arp 299). 
And there are smaller-scale differences: the anti-correlation between
the edges of gaseous and optical features depicted in
Fig.~\ref{fig:slices}.  A similar anti-correlation is observed between
\hi\   and optical shells in shell galaxies (Schiminovich \et\   1994a,b,
1999), many of which are believe to be more evolved merger remnants. 

We have examined a number of possible explanations for these
observations, including dust obscuration, differences in the original
distribution of gas and starlight in the progenitor disks, gas cloud
collisions within the developing tails, ram pressure stripping due to
an extensive hot halo or an expanding superwind, and photoionization
by either the central starburst or evolved sources in the tidal tails
themselves.  However, no one model easily and completely explains the
observations, and it is conceivable that all explanations are playing
a role at some level.

The most likely explanation for
the lack of starlight associated with the outer tidal \hi\
is that such features arise from the
\hi-rich but optically faint outer radii of the progenitor
disks. The relatively short lifetimes of luminous stars and the large
velocity dispersions of less luminous stars, especially with respect
to the gas, will further dilute the luminous content of this material,
and the \hi -to-light ratio of the resulting tidal features will
increase with time (Hibbard \et\   1994).  Gaseous tidal extensions with
very little detectable starlight would seem to be the natural
consequence.  The outer \hi\   rings in NGC 520 and Arp 220 and the
gas-rich outer filament in Arp 299 are all likely to arise from these
gas-rich regions of their progenitor disks. 

For the gas-poor tidal features we suggest that the starburst has
played an important role in shaping the gaseous morphology, either by
sweeping the features clear of gas via a high-pressure expanding
superwind, or by excavating a clear sightline towards the starburst
and allowing ionizing photons to penetrate the tidal regions.  The
primary supporting evidence for this conclusion is rather
circumstantial: the five galaxies with the most striking \hi/optical
displacements (the three systems currently under study here,
and the \hi\  accreting starburst systems M82 and NGC
4631) host massive nuclear starbursts with associated powerful
outflows or superwinds aligned with the direction of the most extreme
\hi/optical displacements.  

NGC 520, Arp 299, and Arp 220 each experienced prograde/polar or
prograde/retrograde encounters.  This relative geometry may be a
pre-requisite for the morphological differences reported here.
Retrograde and polar encounters do not raise extensive tidal tails
(e.g. Barnes 1988), leaving large gaseous disks in the inner
regions. These disks should help collimate and ``mass-load'' the
superwind (Heckman, Lehnert \& Armus 1993; Suchkov \et\ 1996), which
in turn leads to denser and longer-lived winds. Simultaneously, the
combination of opposite spin geometries provides the opportunity for
the tidal tail from the prograde system to rise above the starburst
region in the polar or retrograde system, where it may intersect the
escaping superwind or UV radiation.  If this suggestion is correct,
only systems hosting a galactic superwind and experiencing a
high-inclination encounter geometry should exhibit such extreme
differences between their \hi\ and optical tidal morphologies.

The observations do not allow us to discriminate between either the RPS
or the photoionization models: simple calculations suggest that either is
capable of affecting the diffuse outer gas if the geometry is right. 
There might be some evidence for the effects of an impinging wind on the
outer material in Arp 299 from the increased velocity dispersion at
several points (HY99); however NGC 520 and Arp 220 show no such
signatures.  Photoionization is an attractive solution, as it offers a 
means of explaining the lack of tidal \hi\ found at the base of otherwise 
gas-rich tidal tails in mergers which show no evidence of a superwind 
(e.g. NGC 7252, Arp 105, NGC 4039; see \S~\ref{sec:ionizSB}). 

Since any ionized hydrogen will emit recombination lines,
both explanations can be checked observationally.  The expected emission
measure is given by eqn.~(\ref{eq:EM}), which predicts detectable
features at the column densities of interest.  The morphology of the
ionized gas should reveal the nature of the ionizing source:
photoionized gas should be smoothly distributed, while gas excited by
RPS should be concentrated in dense shocked regions on the edges of the
\hi\   that are being compressed by the superwind, \ie\   on the edges
nearest the wind axis in Figs.~\ref{fig:n520mos}d, \ref{fig:a299mos}d \&
\ref{fig:a220mos}d.  If the gas is clumpy, there may be regions bright
enough to allow line ratios to be measured, which should further aid in
discriminating between photoionization or shock excitation. 

Only two scenarios are offered to explain the small-scale
anti-correlations: dust obscuration and photoionization due to evolved
sources in the optical tails.  Dust obscuration likely affects the
apparent tidal morphologies at the lowest light levels, but we suspect
that the dust content is too low to significantly obscure the brighter 
tidal features. However, if the tidal tails are highly ionized, with
the neutral gas representing only a small fraction of the total hydrogen
column density, it is possible that we are grossly underestimating the
expected amount of absorption.  This question can be investigated
directly with submm imaging of the cold dust in tidal tails. 

The other possibility is that the UV flux from evolved sources in the
optical tails is responsible for ionizing nearby diffuse outer \hi.  A
simple calculation suggest that the tidal tail in Arp 220 is bright
enough to ionize nearby \hi, but the expected ionization flux from the
optical tails in NGC 520 and Arp 299 is too low to explain the
observed differences, unless significant star formation continued
within these features after their tidal ejection.  If this is indeed
the case, then the regions where the neutral gas column density drops
rapidly (see Fig.\ref{fig:slices}) should contain ionized gas which
would emit recombination radiation.  The expected levels of emission
should be observable with deep imaging techniques (see above). This
situation requires that the gas and stellar features are physically
close, and not just close in projection, which can tested with
detailed numerical simulations.

\acknowledgments

We would like to thank Lee Armus Tim Heckman for sharing of
unpublished results, and Rhodri Evans, Jacqueline van Gorkom, 
Dave Schiminovich, and Josh Barnes for useful discussions. We thank
the referee, Chris Mihos, for a thorough and useful report.


\clearpage

\begin{deluxetable}{lrlllc}
\tablecaption{Effects of Ram Pressure and Ionization on Tidal Material}
\tablehead{
\colhead{Parameter} &
\colhead{Units} &
\colhead{NGC 520} &
\colhead{Arp 299} &
\colhead{Arp 220} &
\colhead{notes}
}\startdata
$V_{hel}$ & (km s$^{-1}$) & 2260 & 3080 & 5400 & \\
$D$       & (Mpc) & 30 & 48 & 79 & $a$ \\
$L_{IR}$  & (L$_\odot$) & $7.6\times 10^{10}$ & 
$8.1\times 10^{11}$ & $1.5\times 10^{12}$ & $b$ \\
$N_{HI,max}$ & (cm$^{-2}$) & $5\times 10^{20}$ &
$4\times 10^{20}$ & $4\times 10^{20}$ & $c$ \\
$\sigma_{HI,max}$ & (km s$^{-1}$) & 16 & 20 & 40 & $c$ \\
$r$ & (kpc) & 25 & 70 & 30 & $d$ \\
\\
$R_{RPS}$ & (kpc) & 50 & 140 &  95 & $e$ \\
$R_{ionize}$ & (kpc) & 25 & 95 & 130 & $f$ \\
\enddata
\tablenotetext{a}{Adopting the distances from Sanders, Scoville
\& Soifer (1991), which assumes $H_o$=75 km s$^{-1}$ Mpc$^{-1}$ and
the Virgocentric flow model of Aaronson \et\  (1982).}
\tablenotetext{b}{8---1000 $\mu$m luminosity, from Sanders, Scoville, 
\& Soifer (1991).}
\tablenotetext{c}{Maximum values of gas column density and line-of-sight 
velocity dispersion observed within the tidal features, taken from HvG96 
(NGC 520), HY99 (Arp 299) and Yun \& Hibbard 1999a (Arp 220).}
\tablenotetext{d}{The projected radius of ``missing'' \hi.} 
\tablenotetext{e}{Calculated via eqn.~(\ref{eq:Rrps}), assuming the values 
given in this table and $dL$=10 kpc.}
\tablenotetext{f}{Calculated via eqn.~(\ref{eq:Rion}), assuming the values 
given in this table and $f_{esc}(\Omega_{wind})$=0.1 and $dL$=10 kpc.}
\label{tab:results}
\end{deluxetable}
\clearpage
\end{document}